\documentstyle[12pt]{article}
\def\met{\mbox{${\hbox{$E$\kern-0.6em\lower-.1ex\hbox{/}}}_T$}} 
\def\D0{D\O}                            
\def\etal{{\sl et al.}}                 

\hsize=\textwidth
\vsize=\textheight
\begin{document}
\vspace*{0.50in}

\begin{center}
{\Large\bf
Combined Limits on First Generation \\ Leptoquarks from the CDF and D\O 
\\[0.075in] Experiments}

\vspace*{0.3in}

{\large
Leptoquark Limit Combination Working Group\footnote{
Leptoquark Limit Combination Working Group:\par\noindent
Carla Grosso-Pilcher, University of Chicago, Chicago, IL 60637, 
carla@uccdf.uchicago.edu\par\noindent
Greg Landsberg, Brown University, Providence, RI 02912, 
landsberg@hep.brown.edu\par\noindent
Marc Paterno, Rochester University, Rochester, NY 14627, 
paterno@fnal.gov}\\[0.075in]
(for the CDF and D\O\ Collaborations)}

\end{center}

\vspace*{0.1in}

\begin{abstract}
We have  combined recently  published  leptoquark results  from the CDF and
D\O\   Collaborations  which  yielded  95\%  CL lower  limits on  the first
generation  scalar leptoquark  mass of  213~GeV and 225  GeV, respectively,
under assumption of  100\% branching fraction of  the leptoquark decay into
the $eq$ channel. The  combined limit from the  two experiments is 242~GeV.
This is the most stringent limit on  the first generation scalar leptoquark
mass to date.
\end{abstract}

\section{Introduction}

Recently, the  D\O\ and CDF  Collaborations  at the  Fermilab Tevatron have
both   published~\cite{D0-lq,CDF-lq} limits  on the pair  production of the
first generation scalar leptoquarks that ruled out an interpretation of the
HERA      high-$Q^2$   event    excess    reported  by  the   H1  and  ZEUS
Collaborations~\cite{Hera-H1,Hera-ZEUS}   as  an  $s$-channel production of
leptoquarks  with 100\%  branching fraction  to the charged  lepton channel
($eq$). D\O\ set a 95\% confidence level (CL) lower limit of 225~GeV on the
mass of such a leptoquark; the analogous CDF limit is 213~GeV.

In  this  paper  we  discuss  a   combination of  the   results of  the two
experiments, using both  Bayesian and  traditional  frequentist approaches,
that results in a  tighter limit of  242 GeV. This is  the most restrictive
limit on the leptoquark mass to date.

\section{Individual results}

The    experimental  results  from  both   experiments  are   summarized in
Table~\ref{table:cdfd0}.  In the  region of interest  (leptoquark mass $M >
200$~GeV) neither experiment observes any candidate events. 

\begin{table}[htb]
\begin{center}
\caption{Individual results from the CDF and D\O\ experiments}
\vskip 0.1in
\label{table:cdfd0}
\begin{tabular}{||l|c|c||}   
\hline\hline  
Quantity & CDF &  D\O\ \\ 
\hline
Number of candidates & 0  & 0 \\ 
\hline  
Background & N/A & $0.44 \pm 0.06$ \\  
\hline   
Efficiency &  28\% $\pm$ 11\% &   $36\%-39\% \pm 12\%$ ($M  >   220$~GeV)\\ 
\hline
Integrated luminosity & 110 $\pm$ 8 pb$^{-1}$ & 123 $\pm$ 6.5 pb$^{-1}$ \\ 
\hline\hline
\end{tabular} 
\end{center}
\end{table}

To obtain a combined limit, we need  to understand the correlations between
the two experiments. We  have considered a number  of systematic errors for
different     parameters  used  in  the    calculation of  the   integrated
luminosities,  efficiencies, and  backgrounds for each  of the experiments.
Central values of  these parameters  and fractional  errors in each of them
are summarized in  Table~\ref{table:combined}. Let us discuss some of these
errors in more details. 

CDF determines the  integrated luminosity based  on its own measurements of
the inelastic, single  and double diffractive  cross sections. The D\O\
calculation of the integrated  luminosity  is  based  on both the
CDF~\cite{CDF}  and   E710~\cite{E710} inelastic,  single  diffractive, and
double diffractive cross  section measurements.  Since the two measurements
of the inelastic cross section  differ by nearly two standard deviations, a
$\chi^2$-based factor of 1.85 is  used to scale the errors in the inelastic
cross sections of the two  experiments (see \cite{d0-lum} for details). The
single and  double  diffractive  cross section  measurements  are in a good
agreement   and  therefore  no  scaling  of  the  errors  quoted by  either
experiment was done for these cross sections. For simplicity of calculation
we use  here  just the  average  of  single and  double   diffractive cross
sections for both the CDF and D\O\ luminosity calculation (CDF in fact uses
just their own  measurement but numerically it  does not affect the results
at all).  We further  neglect  the error  on the  double  diffractive cross
section  since it is  small  compared to  the errors  in the  inelastic and
single  diffraction cross  sections. The  CDF- and  D\O-specific luminosity
errors are then  calculated by  subtracting in  quadrature the error due to
the cross section  measurements (based on each  experiment's approach) from
the total quoted errors of 7.2\% (CDF) and 5.3\% (D\O). 

The other source  of common  systematics is the error  in efficiency due to
the MC  modelling of  the  signal,  dominated by the   uncertainties due to
parton  distribution functions and  gluon  radiation. This fractional
10\%  error  is  assumed  to be   completely   correlated  between  the two
experiments and folded in the  efficiency for each experiment as a $1.0 \pm
0.1$ factor. Then the CDF- and  D\O-specific efficiency errors are obtained
by subtracting the common 10\% error  in quadrature from the overall quoted
11\%   (CDF)  and   12\%  (D\O)    efficiency   errors,   which   gives the
experiment-specific errors of: 4.9\% (CDF) and 6.5\% (D\O).

\begin{table}[hbt]
\begin{center}
\caption{Integrated luminosity, efficiency and background for CDF and D\O\
leptoquark searches.}
\vskip 0.1in
\begin{tabular} { ||l| c| c||}
\hline\hline
Parameter & Center value & Fractional uncertainty \\
\hline\hline
CDF inelastic c.s.              & 60.33 pb      & 2.3\%  \\
\hline
E710 inelastic c.s.             & 55.5 pb       & 4.0\%  \\
\hline
Average single diffractive c.s. & 9.54 pb       & 4.5\%  \\
\hline
CDF-specific luminosity error   & 110 pb$^{-1}$ & 6.6\%  \\
\hline
D\O-specific luminosity error   & 123 pb$^{-1}$ & 2.6\%  \\
\hline
Common MC modelling factor      & 1.00          & 10.0\% \\
\hline
CDF-specific efficiency error   & 0.28          & 4.9\%  \\
\hline
D\O-specific efficiency error   & 0.38          & 6.5\%  \\
\hline
CDF background                  & N/A           & N/A    \\
\hline
D\O\ background                 & 0.44          & 13.6\% \\
\hline\hline
\end{tabular}
\label{table:combined}
\end{center}
\end{table}

\section{Bayesian Approach}

In the Bayesian aproach we  will define  probability
functions for  each  experiment in such a  way that they  take into account
correlated and uncorrelated uncertainties. We assume Gaussian errors in all
the parameters. 

We start  by  determining the  95\%  CL upper limits  from each  experiment
independently, i.e. reproducing published numbers.

Since neither  experiment  observed any  candidate events  in the region of
interest, the likelihood  function of each  measurement is given by Poisson
distribution with given expected signal and background means:
\begin{equation}
    P(n=0\;|\; S({\cal L},\varepsilon),B,I) = e^{-(S+B)}, \label{eq:prob}
\end{equation}
where   $S({\cal   L},\varepsilon)$  is the   expected  signal for  a given
integrated luminosity  ${\cal L}$ and efficiency  $\varepsilon$, $B$ is the
expected  background,  and $I$  represents other  prior  information. For a
given cross section of the leptoquark pair production $\sigma$ we have:
\begin{equation}
    S({\cal L},\varepsilon) = \sigma\varepsilon{\cal L}. \label{eq:signal}
\end{equation}
We can  now apply  Bayes'  theorem and  write down  the  expression for the
posterior probability for the cross section $\sigma$, given the observation
of zero events in the data:
\begin{equation}
    p(\sigma \;|\; n=0, I) = \frac{\int d\varepsilon d{\cal L} dB P(n=0\; |
    \sigma{\cal L}\varepsilon,B,I) p(\varepsilon|I) p({\cal L}|I) p(B|I)
    p(\sigma|I)}{\int d\sigma d\varepsilon d{\cal L} 
    p(\varepsilon|I) p({\cal L}|I) p(B|I) p(\sigma|I)}, \label{eq:posterior}
\end{equation}
where  $p(\varepsilon|I)$, $p({\cal L}|I)$,  $p(B|I)$ are prior probability
densities for  efficiency,  integrated luminosity and  backgrounds, and are
Gaussian by assumption. Finally,  $p(\sigma|I)$ is the prior for the signal
cross section, and since the most basic assumption about the signal is that
it cannot be negative,  but otherwise can be anything, a natural choice for
the signal prior  is $p(\sigma|I) =  \theta(\sigma)$,  where $\theta(x)$ is
$\theta$-function, defined as $0$ for $x<0$ and 1 for $x \ge 0$.

An efficient way to calculate the integral (\ref{eq:posterior}) is to use a
Monte Carlo  (MC) integration  by generating  random values  of ${\cal L}$,
$\varepsilon$ an $B$  according to their Gaussian  priors. The value of the
integral   is   simply  the   average  value  of    $p(n=0\;|   \sigma{\cal
L}\varepsilon,B,I)$   obtained  in the  series of  the MC  trials, since by
definition probability density functions are normalized to unity.

We then  vary the  input  value of  $\sigma$  for the  MC trials  to obtain
$P(\sigma\;|\;n=0,I)$  in the  entire range:  $0 \le \sigma  < \infty$. The
upper 95\% confidence  level limit on signal  cross section, $\sigma^{95}$,
can then be obtained by solving the following integral equation:
\begin{equation}
    \int_0^{\sigma^{95}}       P(\sigma\;|\;   n=0,I)d\sigma  =  0.95
    \int_0^\infty p(\sigma\;|\; n=0,I)d\sigma. \label{eq:s95}
\end{equation}
(Here  we have  to  normalize  the  posterior   probability to  unity since
$p(\sigma|I) = \theta(\sigma)$ is not properly normalized.)

From the Eq.  (\ref{eq:prob}) it is  natural to expect  that $p(\sigma\;|\;
n=0,I)$ can  be  parameterized as $A   e^{-a\sigma}$. In this  case equation
(\ref{eq:s95}) transforms into:
$$
    \exp(-a\sigma^{95}) = 1 - 0.95 = 0.05,
$$
which can be easily solved:
\begin{equation}
    \sigma^{95} = \frac{1}{a}\ln 20 = 3.00/a. \label{eq:s95exp}
\end{equation}

We  can  now  apply  this  approach    independently to  the  CDF  and D\O\
measurements. The only  subtlety here is that the  efficiency $\varepsilon$
generally depends on the leptoquark mass, and thus far we have not made any
connections between the  mass and the cross  section. One can obtain a more
complex two-dimensional function $p(\sigma,M\;|\; n=0,I)$ and then obtain a
cross section limit for any given leptoquark mass. We however, will use the
fact that for the  leptoquark masses  above 200 GeV the  efficiency changes
very slowly,  and we simply  use the  efficiency measured  at the published
value of  the mass  limit for  each  experiment in  order to  reproduce the
results. When calculating the limits  for each particular experiment, we do
not care about correlated and uncorrelated errors, and therefore can simply
use overall  uncertainties on  ${\cal L}$,  $\varepsilon$  and $B$ for each
experiment. In the case of CDF,  there was no background estimate made, and
therefore no background  subtraction was used  when obtaining cross section
limits. This fact,  however, does not influence  the limit since it is well
known~\cite{PDG} that  for the case of zero  observed candidate events, the
limit on the signal does  not depend on the  actual value of the background
or its  uncertainty. We  therefore  assign  background $B =  0.00 \pm 0.00$
events in the CDF case.

\begin{figure}
\vspace*{3.0in}                      
\includegraphics{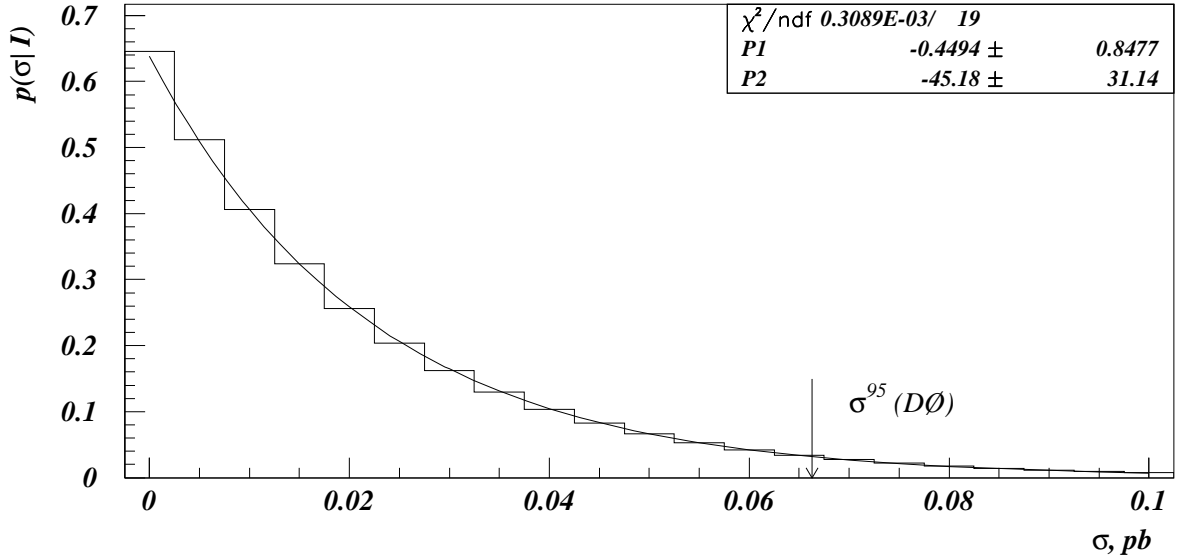}
\caption{Posterior  probability  $p(\sigma\;|n=0,I)$ for  D\O\ experiment and
the corresponding  95\% CL cross section upper   limit. The line shows a fit
to the   exponential  function. The  arrow  indicates  95\% CL  cross
section upper  limit.}
\label{fig:d0}
\end{figure}

The   estimates of  ${\cal  L}$,   $\varepsilon$  and $B$  for  each of the
experiments are summarized in Table~\ref{table:separate}.
\begin{table}[hbt]
\begin{center}
\caption{Integrated luminosity, efficiency and background for CDF and D\O\
leptoquark searches.}
\vskip 0.1in
\begin{tabular} { ||l| c| c||}
\hline\hline
Parameter & CDF & D\O\ \\
\hline\hline
${\cal L}$      & $110 \pm 8$ pb$^{-1}$ & $123 \pm 7$ pb$^{-1}$ \\
\hline
$\varepsilon$   & $(28.0 \pm 3.1)$\%    & $(38.0 \pm 4.6)$\%    \\
\hline
$B$             &       N/A             & $0.44 \pm 0.06$       \\
\hline\hline
\end{tabular}
\label{table:separate}
\end{center}
\end{table}
A   posterior   probability   function for  D\O\   experiment  is  shown in
Fig.~\ref{fig:d0},      together  with an   exponential  fit.  The  fit has
essentially     $\chi^2=0$,  so  indeed  the    assumption of   exponential
approximation   works  well.  Using    Eq.~(\ref{eq:s95exp}) we  obtain the
following 95\% CL upper limits on the production cross sections:
\begin{eqnarray}
    \sigma^{95}_{\rm CDF} & = & 0.0993~\mbox{pb} \\
    \sigma^{95}_{\rm D\O} & = & 0.0663~\mbox{pb},
\end{eqnarray}
which can be translated into the  lower mass limits on the first generation
scalar     leptoquarks  using     parameterization  of  the   lower  band of
next-to-leading order cross section~\cite{Kramer}:
\begin{eqnarray}
    M & > & 213.5~\mbox{GeV (CDF)} \\
    M & > & 225.8~\mbox{GeV (D\O)},
\end{eqnarray}
in agreement with the published values~\cite{CDF-lq,D0-lq}.

Having  reproduced the  individual results  of each  experiment, we can now
combine them by  breaking down the  errors of each of  the experiments into
several   pieces and  by  using one  and the  same  random  values  for the
correlated  uncertainties, and  different values for  the uncorrelated ones
during the MC integration.

For the  combined  results  from the  two  experiments  we again  have zero
candidates    observed, so  the  only   required   modification to  the Eq.
(\ref{eq:prob}) is that the signal expectation is:
\begin{equation}
    S = \sigma(\varepsilon_1{\cal L}_1 + \varepsilon_2{\cal L}_2),
\end{equation}
where   indices 1  and 2   correspond  to the  CDF  and D\O\   experiments,
respectively. The  overall background expectation  is still the same, since
we use no background for  the CDF case. The rest  of the formalism does not
change, except that now  we integrate not over  $d{\cal L}d\varepsilon dB$,
but over  all the   individual  Gaussian  uncertainties  used in  the joint
analysis (see Table~\ref{table:combined}).

\begin{figure}
\vspace*{3.0in}                      
\includegraphics{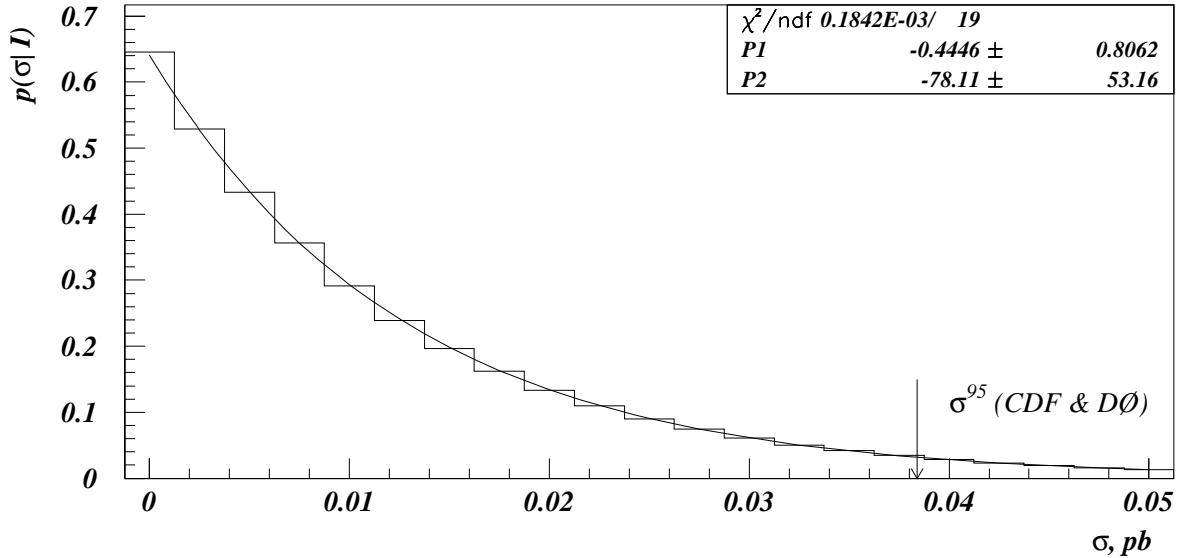}
\caption{Posterior probability  $p(\sigma\;|n=0,I)$ for combined CDF and D\O\
experiments and  the corresponding  upper 95\% CL cross  section limit. The
line shows a fit to the  exponential function.  The arrow indicates 95\% CL
upper cross section limit.}
\label{fig:combined}
\end{figure}

After performing  the MC integration  we obtain the  posterior  probability
function     for   the  CDF   and   D\O\      measurements,  as    shown in
Fig.~\ref{fig:combined},   which is well  described by an  exponential. The
following 95\% CL limits on the production cross section and the leptoquark
mass are obtained:
\begin{eqnarray}
    \sigma & < & 0.0383~\mbox{pb} \label{eq:cs-comb}\\
    M & > & 242.4~\mbox{GeV}.
\end{eqnarray}
This is the final result of the Bayesian combined analysis.

\section{Frequentist Analysis}

In the classical or  frequentist approach, the  upper limit at a confidence
level  $\alpha$ is  defined such  that, when the  experiment  is repeated a
large number  of times, the  fraction of  experiments that  observe a value
smaller or equal to the  measured one is less or  equal to $\alpha$, if the
true value of the quantity is larger than the upper limit. If the number of
observed  events is  small they will  be  distributed  according to Poisson
statistics, $P(n;\mu) =  \frac{e^{-\mu}\mu^n}{n!}$, where $n$ is the number
of observed  events and $\mu$ =  ${\cal L} \sigma  \varepsilon$ is the most
probable   number  of  events   produced.  ${\cal  L}$  is  the  integrated
luminosity, $\varepsilon$ is the efficiency, and $\sigma$ is the production
cross section. If we have two measurements of the  same quantity, the total
likelihood of observing $n_1$ and $n_2$ events from the same source is:
\begin{equation}
      P(n_1,n_2\;|\;\mu) = P(n_1\;|\;\mu)P(n_2\;|\;\mu).
\end{equation} 

To account for  systematic uncertainties on the  efficiency and luminosity,
the Poisson probability is convoluted with the probability distributions of
$\varepsilon$ and ${\cal L}$, assumed to be Gaussians. Then the probability
of observing a number of events $n_i$ in experiment $i$ is:
\begin{equation}
{\it P(n_i\;|\;{\cal L}\varepsilon_i \mu)} = 
\int_0^1{\it P(n_i\;|\;{\cal L}\varepsilon_i' \mu)} 
 {\it G(\varepsilon_i';\varepsilon_i, \sigma_i)}d\varepsilon_i',
\end{equation}
where  ${\it  G(\varepsilon_i'\;|\;\varepsilon_i, \sigma_i)}$ indicates the
sampling   of  the   efficiency  with a   Gaussian    distribution  of mean
$\varepsilon_i$ and  sigma  $\sigma_i$.

In  combining two  experiments the   uncertainties in the  efficiencies are
decomposed in two parts: those  correlated between the two experiments, and
the   uncorrelated ones,  as  indicated  above.  The  total  likelihood for
observing a set of events in the two experiments is:
$$
 {\cal L} = \prod_{1}^{2}{\it P(n_i\;|\;{\cal L}_i\varepsilon_i \mu)},
$$
with the correlation between the efficiencies taken into account:
$$
{\it G(\varepsilon'_i;\varepsilon_i, \sigma_i)} = 
{\it G(\varepsilon'^{\rm ~corr}_i\;|\;\varepsilon_i^{\rm corr}, 
\sigma_i^{\rm corr})}\cdot {\it G(\varepsilon'^{\rm ~uncorr}_i\;|\;
\varepsilon_i^{\rm uncorr}, \sigma_i^{\rm uncorr})}.
$$
The likelihood is  evaluated with a  Monte Carlo  technique~\cite{cdf1109}.
The mean number  $\mu_i$ of events  for each experiment  is calculated from
the  relative  luminosities and  efficiencies,  with the  latter drawn from
Gaussian      distributions  with  a  common   width  for  the   correlated
uncertainties, and individual widths  for the uncorrelated ones. The number
of  observed events  $n_i'$ for  each  experiment is  drawn  from a Poisson
distribution    with  mean   $\mu_i$.  By   generating a   large  number of
experiments,  we calculate  the fraction  with $n_i'$ less  or equal to the
observed events (zero in this case)  for the different values of $\mu$. The
95\% upper limit corresponds to the  $\mu$ for which this fraction is $\le$
0.05. With the values of the efficiencies discussed above, we obtain a 95\%
cross section limit of  0.0391 pb, which  corresponds to a lower mass limit
of 241.7  GeV for scalar   leptoquarks. This  calculation does not take into
account backgrounds in each  experiment, which in the case of zero observed
events is equivalent to  a proper background  subtraction technique. Limits
obtained by the frequentist method  are in a good agreement with those from
the Bayesian prescription.

\section{Conclusions}

\begin{figure}
\vspace*{6.7in}                      
\includegraphics{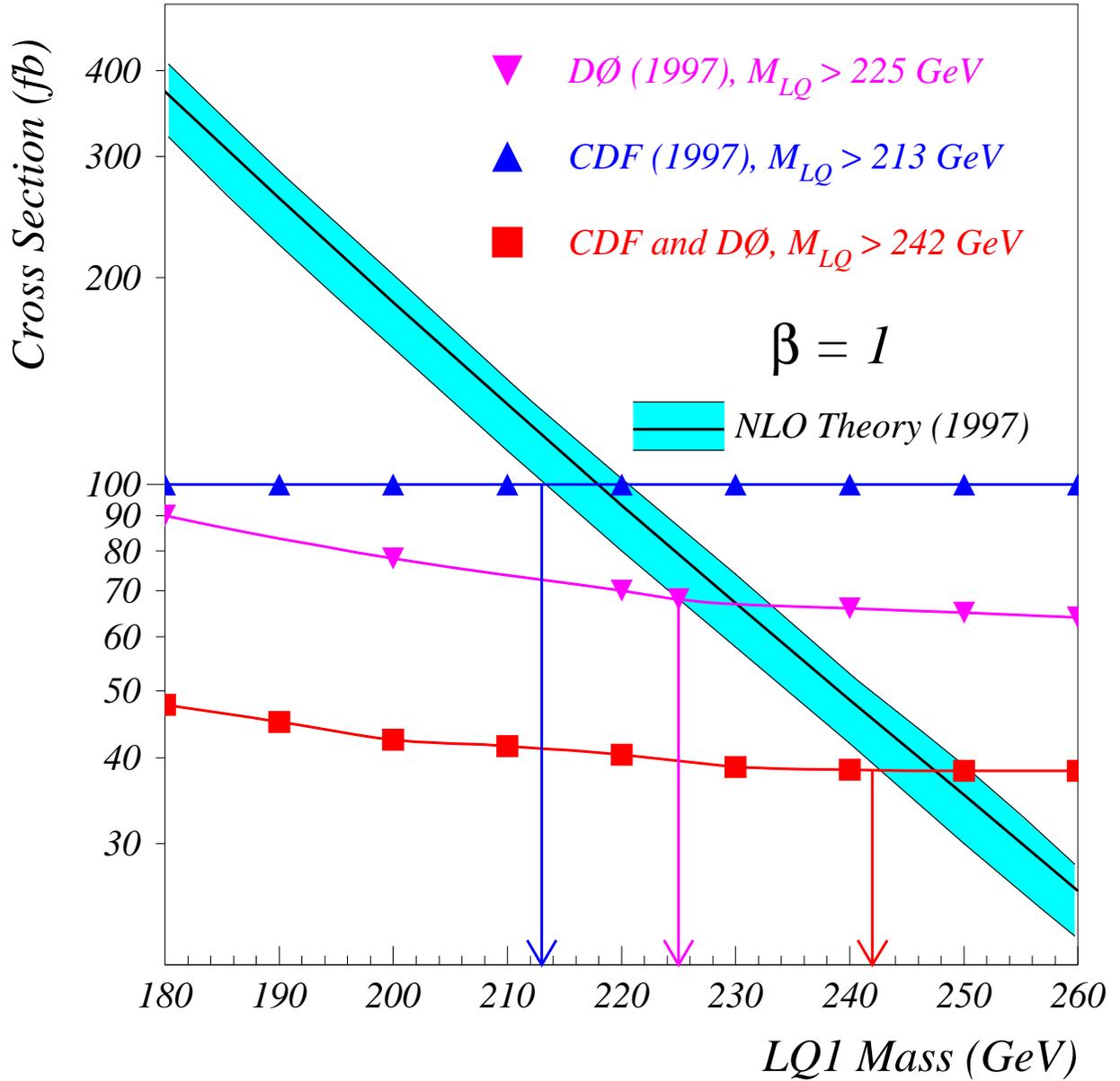}
\caption{95\% CL upper cross section limits from CDF (triangles), D\O\
(inverted triangles), and combined (squares) leptoquark analyses. The band
shows the NLO theoretical cross section; arrows correspond to the
respective 95\% CL lower mass limits.}
\label{fig:comb}
\end{figure}

We performed  proper statistical  combinations of the  D\O~\cite{D0-lq} and
CDF~\cite{CDF-lq} limits on the  first generation scalar leptoquarks, using
both   Bayesian  and   frequentist   approaches,  with   common  systematic
uncertainties taken into account.  The two methods are in a good agreement.
Our  combined  upper  95\% CL  cross   section  limit for   leptoquark pair
production is 38~fb, which corresponds to a lower LQ mass limit of 242~GeV,
based on the  lower band of the NLO   calculations~\cite{Kramer}. The cross
section limits from both  experiments, as well as  the combined result, are
shown in  Fig.~\ref{fig:comb},  together with the  theoretical predictions.
This is the  most stringent  limit on the  mass of first  generation scalar
leptoquarks to date.

\section{Acknowledgements}

We  would like  to  thank all  the CDF  and D\O\  members   involved in the
leptoquark  searches. We  thank the  staffs at  Fermilab and  collaborating
institutions for their contributions  to this work, and acknowledge support
from the Department of Energy and National Science Foundation (U.S.A.), the
Italian  Istituto Nazionale  di Fisica  Nucleare; the  Ministry of Science,
Culture,  and  Education of  Japan; the  Natural  Sciences and  Engineering
Research Council of Canada; the National Science Council of the Republic of
China; Commissariat \` a  L'Energie Atomique  (France), State Committee for
Science and Technology  and Ministry for Atomic  Energy (Russia), CAPES and
CNPq  (Brazil),  Departments of  Atomic  Energy and  Science  and Education
(India),  Colciencias (Colombia),  CONACyT (Mexico),  Ministry of Education
and KOSEF  (Korea),  CONICET and  UBACyT  (Argentina), and  the A. P. Sloan
Foundation.

\end{document}